\documentclass[%
 aip,
 jmp,%
 amsmath,amssymb,
 reprint,
 ]{revtex4-1}

\usepackage{graphicx}
\usepackage{dcolumn}
\usepackage{bm}%
\usepackage[colorlinks=true,bookmarks=false,citecolor=blue,urlcolor=blue]{hyperref}
\usepackage{upgreek}

\begin{document}

\preprint{AIP/123-QED}

\title{Twin photon pairs in a high-Q silicon microresonator} 

\author{Steven Rogers}

\author{Xiyuan Lu}
\affiliation{Department of Physics and Astronomy, University of Rochester, Rochester, New York 14627, USA}

\author{Wei C. Jiang}
\affiliation{Institute of Optics, University of Rochester, Rochester, New York 14627, USA}

\author{Qiang Lin}
\email[Electronic mail: ]{qiang.lin@rochester.edu}
\affiliation{Institute of Optics, University of Rochester, Rochester, New York 14627, USA}
\affiliation{Department of Electrical and Computer Engineering, University of Rochester, Rochester, New York 14627, USA}

\date{\today}

\begin{abstract}
We report the generation of high-purity twin photon pairs through cavity-enhanced non-degenerate four-wave mixing (FWM) in a high-Q silicon microdisk resonator. Twin photon pairs are created within the same cavity mode and are consequently expected to be identical in all degrees of freedom. The device is able to produce twin photons at telecommunication wavelengths with a pair generation rate as large as $(3.96\pm0.03)\times10^5 ~{\rm pairs/s}$, within a narrow bandwidth of 0.72 GHz. A coincidence-to-accidental ratio of $660\pm62$ was measured, the highest value reported to date for twin photon pairs, at a pair generation rate of $(2.47\pm0.04)\times10^4 ~{\rm pairs/s}$. Through careful engineering of the dispersion matching window, we have reduced the ratio of photons resulting from degenerate FWM to non-degenerate FWM to less than 0.15. 
\end{abstract}                                                                                                  


\maketitle

High-purity indistinguishable photons underlie many quantum optical phenomena \cite{MandelBook,Ou07,Pan12} and are essential for linear optical quantum computing. \cite{Kok07,Knill00} Their quantum interference forms the foundation of a vast variety of quantum photonic functionalities, such as quantum simulation and quantum optical computing, which have been recently demonstrated on chip-scale devices. \cite{OBrien09,Guzik12,Bose12,Spring13,Tillmann13,Crespi2013,Metcalf14} The importance of high-purity single-mode indistinguishable photons to integrated quantum photonics has excited significant interest in the past few years to develop optical fiber and chip-scale sources for their generation. \cite{Fan05,Chen06,Guo14,He14}

\begin{figure}[b]
\setlength{\belowcaptionskip}{-7pt}
\includegraphics[width=1\columnwidth]{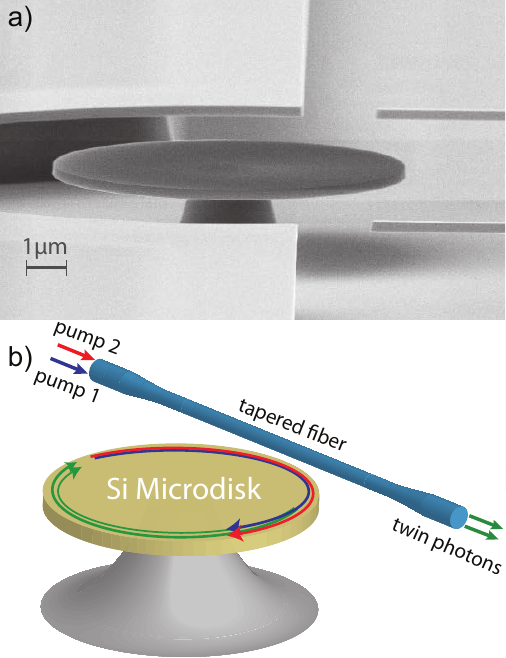}
\caption{\label{Fig1} (a) A scanning electron microscopic (SEM) image of the silicon microdisk sitting on a silica pedestal. (b) A schematic illustrating the device, tapered fiber, and non-degenerate FWM process resulting in twin photon pairs.}
\end{figure}

Here we report the generation of high-purity twin photon pairs on a silicon chip. By taking advantage of dramatic cavity-enhanced non-degenerate four-wave mixing (FWM) in a high-Q silicon microdisk resonator, we are able to produce twin photon pairs in a single temporal and spatial mode within a very narrow bandwidth of 0.72 GHz, with a pair generation rate of $(3.96\pm0.03)\times10^5 ~{\rm pairs/s}$ and high photon-pair correlation, indicated by a coincidence-to-accidental ratio (CAR) as large as $660\pm62$ (at a pair generation rate of $(2.47\pm0.04)\times10^4 ~{\rm pairs/s}$), which is significantly higher than other approaches reported to date.\cite{Fan05,Chen06,Guo14,He14} The demonstrated device exhibits great potential for on-chip quantum photonic applications on the CMOS compatible silicon-on-insulator platform.

\begin{figure}[t!]
\begin{center}
\setlength{\belowcaptionskip}{-7pt}
\includegraphics[width=1\columnwidth]{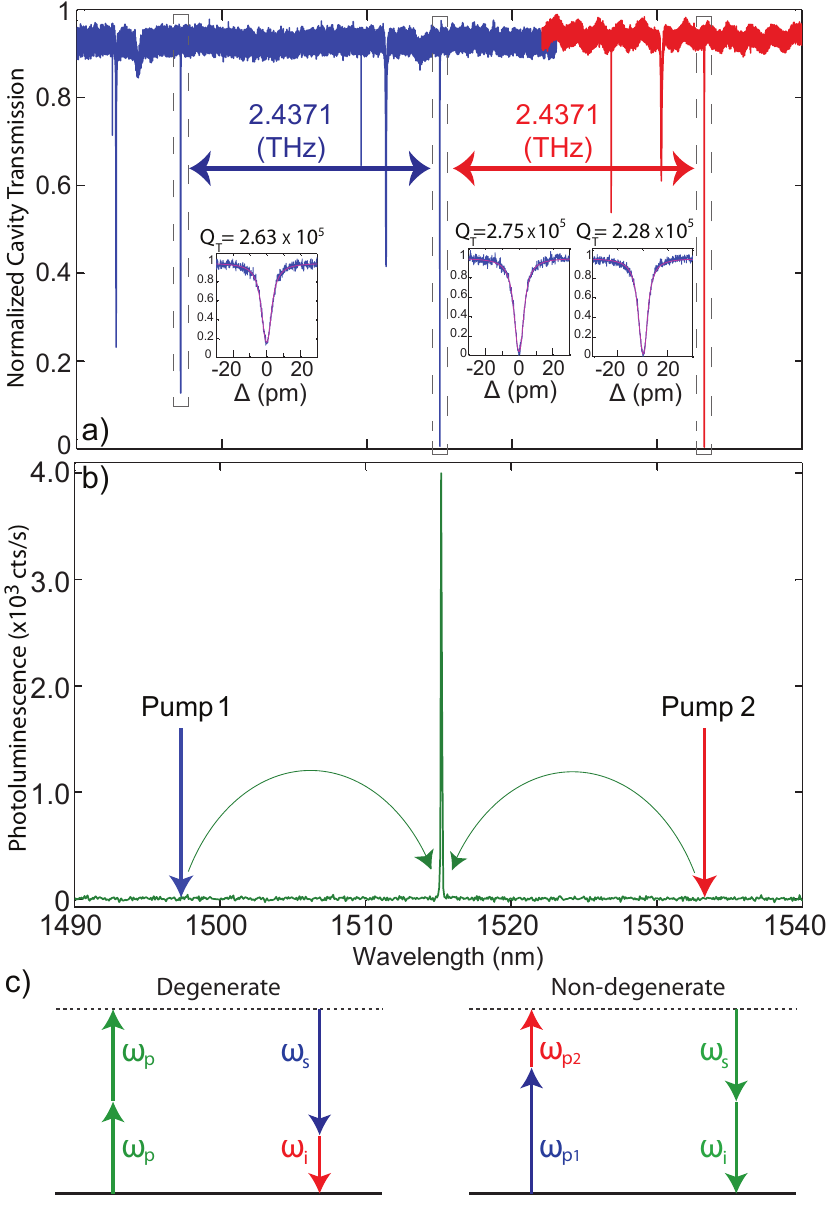}
\caption{\label{Fig2} (a) A trace of the background normalized cavity transmission. The employed mode family is indicated by dashed boxes, with fits of the loaded optical Q shown in the corresponding insets. $\Delta$ = Detuning. The calibrated free spectral range measurements are depicted between adjacent cavity modes. (b) The photoluminescence spectrum as a result of simultaneously pumping at the 1496.8 nm and 1534.3 nm cavity modes. (c) Energy diagrams depicting degenerate/non-degenerate FWM.}
\end{center}
\end{figure}

Figure \ref{Fig1}(a) shows a scanning electron microscopic (SEM) image of the device. The silicon microdisk has a diameter of approximately 9~$\upmu$m and a thickness of 260 nm. A schematic of the pair generation process can be seen in Fig.~\ref{Fig1}(b). Two pump lasers are evanescently coupled into the cavity, where twin photon pairs are created and then coupled out of the cavity through the same tapered fiber. This ensures that the generated photons occupy a single spatial mode, which is integral to high visibility quantum interference. \cite{Ou07,Rohde05}
  	
Figure \ref{Fig2}(a) shows the measured cavity transmission spectrum of the microresonator. The cavity modes employed for twin photon generation are quasi-transverse-magnetic (quasi-TM) and exhibit high optical Qs $\sim 2.5\times10^5$ for the loaded cavity modes (Fig.~\ref{Fig2}(a), insets). The device dispersion was engineered to achieve an equal spacing among the three modes at 1496.8 nm, 1514.7 nm, and 1534.3 nm, as indicated by the calibrated free spectral range values in Fig.~\ref{Fig2}(a). With frequency matching among the three modes, non-degenerate FWM (Fig.~\ref{Fig2}(c)) can efficiently scatter pairs of pump photons from the 1496.8 nm (pump 1) and 1534.3 nm (pump 2) cavity modes to create bi-photon states at the central mode at 1514.7 nm (Fig.~\ref{Fig2}(b)), with an efficiency scaling as $Q/V$ where $V$ is the effective mode volume (see below). A simulation using the finite element method (FEM) indicates that the device exhibits a small effective mode volume of $\sim10~{\rm \mu m}^3$. Consequently, the high optical Q and small effective mode volume significantly strengthens the nonlinear parametric processes inside the cavity, resulting in efficient generation of twin photon pairs. This is evident in Fig.~\ref{Fig2}(b), where the strong photoluminescence spectrum clearly demonstrates that photon pairs were produced at a single cavity mode with a very clean background.

On the quantum mechanical level, twin photon pairs are produced by annihilating two pump photons of different energy eigenstates, resulting in the creation of two new photons belonging to the same energy eigenstate. This process happens instantaneously inside the cavity, in adherence with energy and momentum conservation. \cite{Boyd08} As a result, the created pair of photons are expected to be identical in all degrees of freedom such as temporal wavepacket, spatial profile, frequency, and polarization, which we term as twin photon pairs. This process can be understood as the time reversal of degenerate FWM, which results in photon pairs with distinguishable frequencies (Fig.~\ref{Fig2}(c)), and has been employed in numerous works relating to the generation of correlated photon pairs in silicon waveguides and resonators.\cite{Sharping06,Clemmen09,Azzini12,Davanco12,Engin13}

Detailed analysis of non-degenerate FWM in a microresonator shows that the expected probability density to emit twin photons from the microresonator into the optical fiber, at times $t_1$ and $t_2$, is given by 
\begin{equation}
    	p_c(t_1,t_2) \approx \frac{\Gamma_e^2}{\Gamma_t^2} (4g^2 N_{p1} N_{p2}) \emph{e}^{-\Gamma_t |t_1-t_2|}, \label{Prob_pair}
\end{equation}
where $\Gamma_e$ and $\Gamma_t$ are the external photon coupling rate and loaded cavity photon decay rate, respectively, of the twin-photon mode at frequency $\omega_0$. $N_{p1}$ and $N_{p2}$ are the intracavity photon numbers of the two pumps at frequencies $\omega_{p1}$ and $\omega_{p2}$, respectively. $g = [{c \eta n_2 \hbar \omega_0\sqrt{\omega_{p1}\omega_{p2}}}]/{({n^2} V)}$ is the vacuum coupling rate for non-degenerate FWM, where $\eta$ is the spatial mode overlap fraction between the three modes, $c$ is the speed of light in vacuum, $n_2$ is the Kerr nonlinear coefficient, and $n$ is the refractive index of silicon. Equation (\ref{Prob_pair}) shows that the twin photons are strongly correlated in time, a clear consequence of the simultaneity of their creation. The linewidth of the twin photons is determined by the photon decay rate of the loaded cavity. Therefore, the twin photons in our device exhibit a narrow linewidth of 0.72~GHz (corresponding to an optical Q of $2.75\times 10^5$. Fig.~\ref{Fig2}(a)). The twin-photon generation rate is given by $R_c = (8g^2 N_{p1} N_{p2}) ({\Gamma_e^2}/{\Gamma_t^3}) $. Consequently, a high $Q/V$ would result in a large twin-photon generation rate and long coherence time.

\begin{figure}[t!]
\begin{center}
\setlength{\belowcaptionskip}{-7pt}
\includegraphics[width=1\columnwidth]{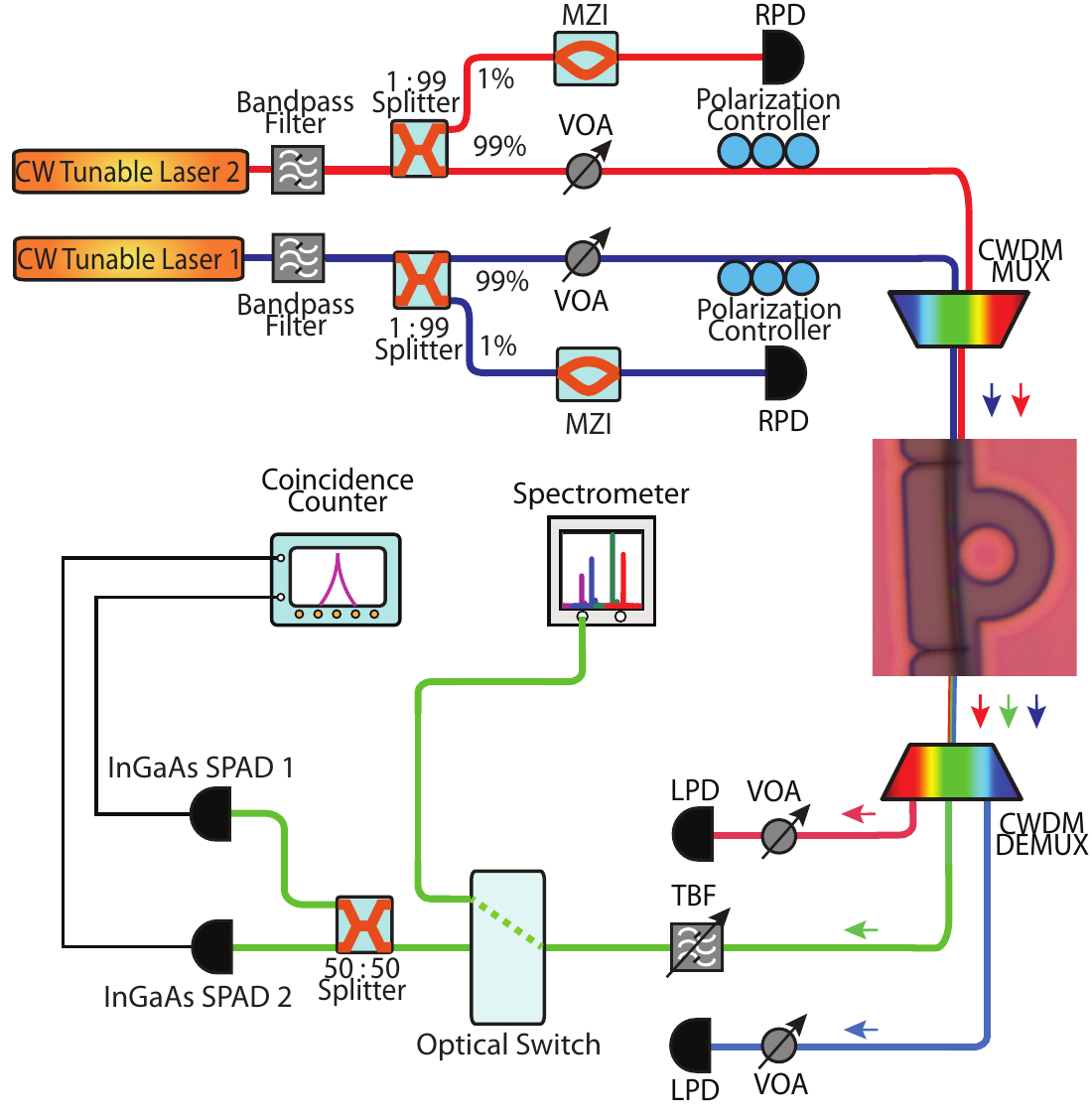}
\caption{\label{Fig3} A schematic of the experimental setup, used to characterize the twin photon source. MZI: Mach-Zehnder Interferometer, VOA: Variable Optical Attenuator, CWDM: Course Wavelength Division Multiplexing, MUX: Multiplexer, DEMUX: De-multiplexer, RPD: Reference Photodetector, LPD: Locking Photodetector, TBF: Tunable bandpass filter, and SPAD: Single Photon Avalanche Photodiode. The inset shows a microscopic image of the device and tapered fiber.}
\end{center}
\end{figure}

To characterize the quality of the twin photon pairs, we built an experimental setup which is schematically depicted in Fig.~\ref{Fig3}. Two tunable continuous-wave (CW) pump lasers were combined in a coarse wavelength division multiplexer (CWDM MUX) and evanescently coupled into the silicon microdisk, through an optical tapered fiber. A microscopic image of the device, with the tapered fiber fixed on adjacent nanoforks, can be seen in the inset of Fig.~\ref{Fig3}. The nanoforks stabilized the transmission signal and allowed for a constant external coupling rate for the duration of the experiment. Twin photon pairs were generated inside the device and delivered out through the same optical fiber. The transmitted composite beam was then passed through a CWDM de-multiplexer (DEMUX) to separate the twin photon mode from the pumps. The twin photon pairs were passed through a narrowband (1.2 nm) tunable bandpass filter and on to an optical switch, which directed the pairs to either a spectrometer or a time-correlated single photon counting (TCSPC) setup. The tunable bandpass filter was used to cut down noise photons generated from spontaneous Raman scattering in the optical fiber. \cite{Jiang12} Note that twin photons share the same frequency and propagate collinearly, so in order to perform coincidence detection, we passed them through a 50/50 beam splitter prior to the coincidence counter. Due to the inherent indistinguishability of the twin photons, coincidence detection results in a measurement of the self correlation of the bi-photon states.

One potential issue of twin photon generation is the contamination from degenerate FWM processes initiated by individual pump modes. Due to the broadband phase matching in micro/nanophotonic devices, the non-degenerate FWM process is generally accompanied with degenerate processes produced by individual pump modes. Consequently, the pump photons may undergo degenerate FWM, which produces photon pairs with one photon located at the twin-photon mode and its temporally correlated partner in a separate cavity mode. These photons are thus uncorrelated with the twin photon pairs and consequently add to the noise background. This issue has presented a major challenge in other schemes for twin photon generation. \cite{Fan05,Chen06,Guo14,He14} Our device, however, allows for the flexible engineering of dispersion which controls the phase matching window \cite{Jiang12,Jiang14} so that it dominantly encompasses the desired cavity modes at 1496.8 nm, 1514.7 nm, and 1534.3 nm (Fig.~\ref{Fig2}(a)), but degrades considerably for the subsequent outer set of cavity modes, thus significantly reducing the efficiency of degenerate FWM from individual pumps. In addition, the tapered-fiber coupling scheme enables flexible control of the external coupling of cavity modes, which in turn can be applied to control the FWM processes involved. As a result, we can optimize the taper-disk spacing to maximize the efficiency of the non-degenerate FWM process while suppressing the degenerate ones. Combining these two advantages, we are able to produce a highly purified non-degenerate FWM process.

\begin{figure}[b]
\begin{center}
\setlength{\belowcaptionskip}{-7pt}
\includegraphics[width=1\columnwidth]{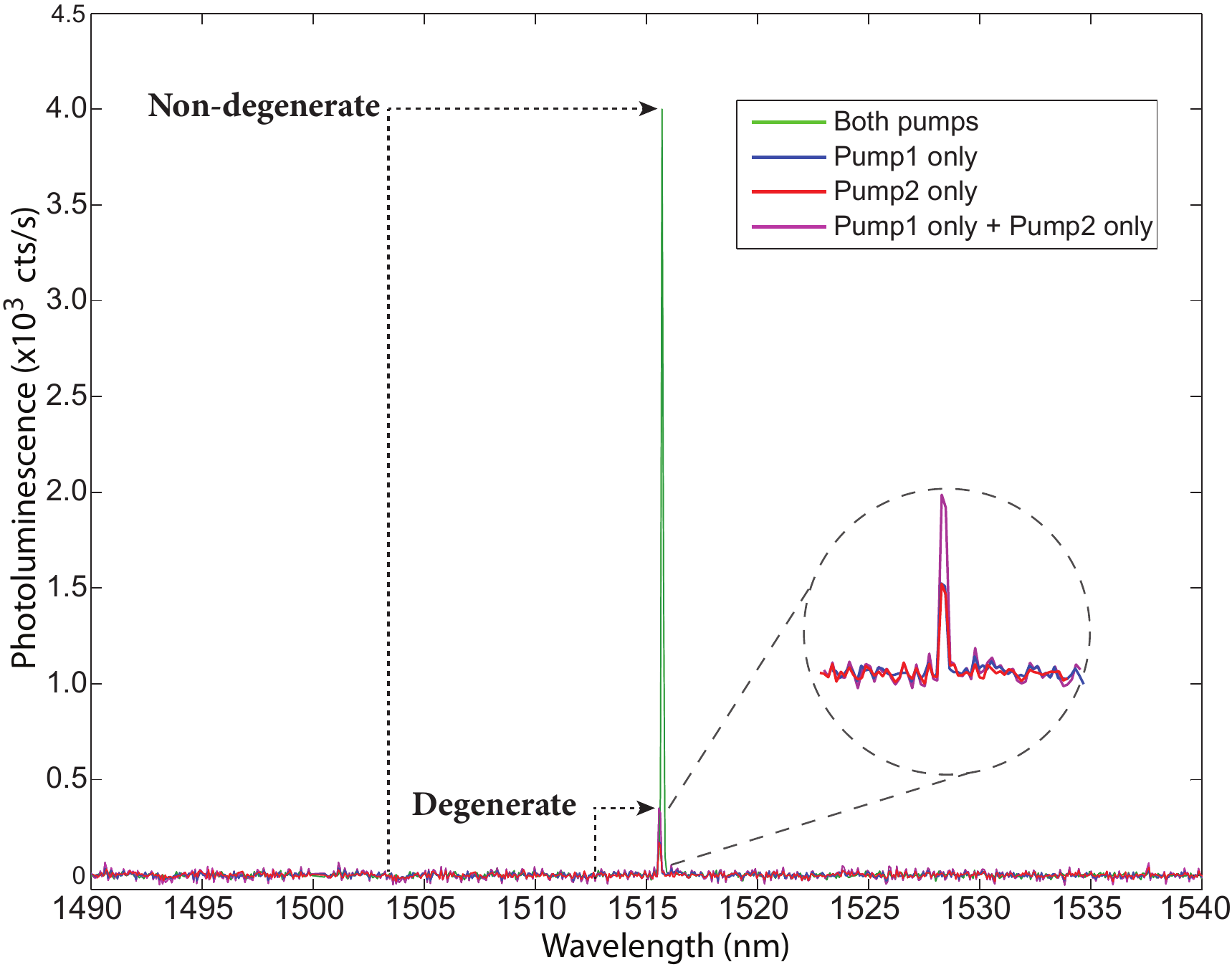}
\caption{\label{Fig4} A plot depicting the relative contributions of degenerate and non-degenerate FWM to the photoluminescence, at a combined dropped optical power of $P_d$ = 37.6 $\mu W$. The red and blue spectra were generated using a single pump. The magenta spectrum results from numerically summing the red and blue spectra. The green spectrum was generated with both pumps on.}
\end{center}
\end{figure}

\begin{figure*}[t!]
\begin{center}
\setlength{\belowcaptionskip}{-7pt}
\includegraphics[scale=1.14]{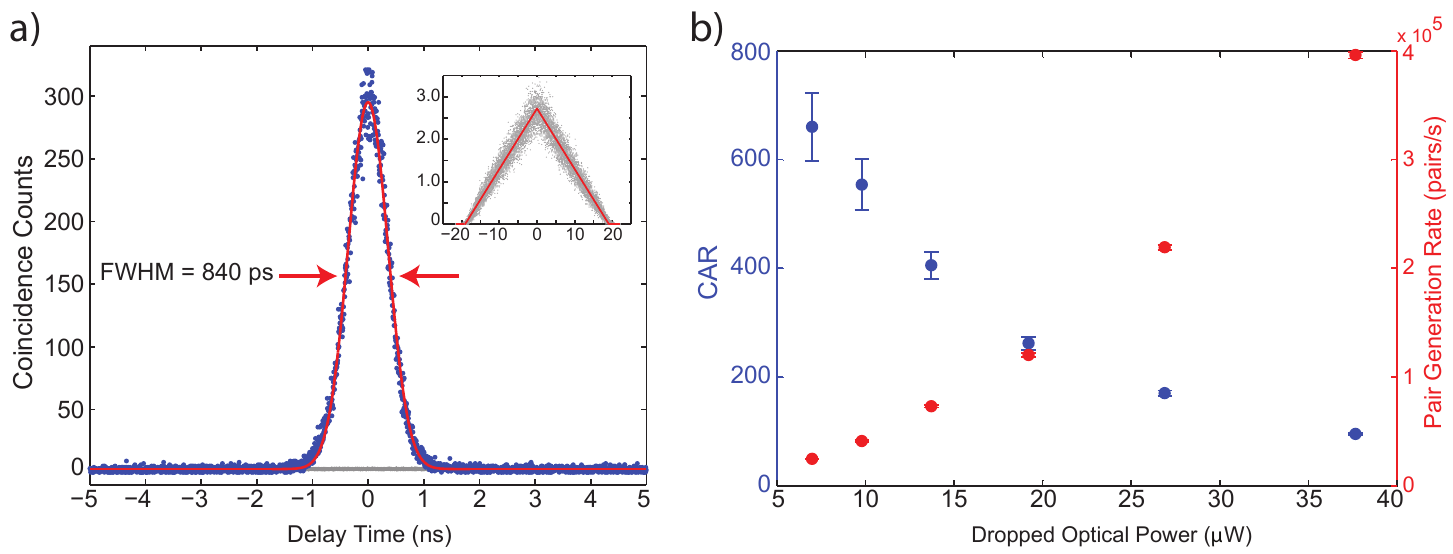}
\caption{\label{Fig5} (a) The coincidence spectrum belonging to the last data point in (b). The coincidence counts are in blue with a Gaussian fit in red. The accidental coincidence counts are shown in gray with a magnified version depicted in the inset. Note that the inset is also plotted as coincidence counts vs. delay time (in ns). (b) A plot of the CAR vs. dropped optical power is shown in blue and corresponds to the left vertical axis. A plot of the pair generation rate vs. dropped optical power is shown in red and corresponds to the right vertical axis.}
\end{center}
\end{figure*}

Figure \ref{Fig4} compares the relative strength of the degenerate and non-degenerate FWM processes. The degenerate FWM contributions were measured in a single pump configuration. The blue photoluminescence spectrum corresponds to pump 1 alone, while the red spectrum corresponds to pump 2 alone, as shown in Fig.~\ref{Fig4}. To indicate their total contribution, the two spectra are numerically summed, as seen in the magenta trace.
Figure \ref{Fig4} shows clearly, that when both pumps are on, the PL spectrum (in green) is significantly larger than those produced by individual pumps. The contribution of degenerate FWM can be quantified as $\dfrac{\sum I_{p1}+\sum I_{p2}}{\sum I_{p1+p2}-(\sum I_{p1}+\sum I_{p2})}$, where $I_j$ ($j=p1, p2, p1+p2$) is the PL intensity of each process. From Fig.~\ref{Fig4}, we estimate that the ratio of degenerate to non-degenerate FWM is $\thicksim$0.14, clearly showing the dominance of non-degenerate FWM over degenerate FWM.

The quantum correlation between twin photon pairs was characterized by coincidence counting measurements. The photons were detected by two InGaAs single photon avalanche photodiodes (SPAD), operated in Geiger mode, with a gating frequency of 2.5 MHz, gate width of 20 ns, dead time of 10 $\mu s$, and quantum efficiency of 15\%. The outputs of the SPADs were transmitted to the TCSPC module for coincidence counting. Figure \ref{Fig5}(a) shows an example of the coincidence spectrum, with a full width at half maximum (FWHM) of $\tau_{\rm FWHM} =$ 840~ps, which is the combined effect of the temporal correlation of the twin photons (see Eq.~(\ref{Prob_pair})) and the instrument response time. The latter was measured to be about 617~ps, primarily resulting from the timing jitters of the two SPADs. To quantify the accidental coincidences, we shifted the time delay by an integer number of the gating period ($T_0=400$~ns) and found the coincidence spectrum. Figure \ref{Fig5}(a) shows an example, in gray, as well as a magnified version in the inset. The difference between the two spectra clearly indicates the high correlation of the twin photons.

The total coincidence counts $C_t$ were obtained by integrating the coincidence spectrum over a certain delay-time window $\Delta \tau$, with a statistical error given by $E_{C_t} = \sqrt{C_t}$. Accordingly, we can obtain the total accidental counts $A_t$ by integrating the accidental coincidence spectrum over the same time window. As $A_t$ is quite small, due to the high purity of the twin photons, we first obtained a series of coincidence spectra with a time delay of $jT_0$ between the two SPADs ($j=1,2,...,40$) and integrated each of them within the same delay-time window $\Delta \tau$ to obtain an ensemble of accidental coincidence counts $A_j$. The average and standard deviation of $A_j$ gave the accidental coincidence counts $A_t$ and its statistical error $E_{A_t}$. Finally, the true coincidence counts were obtained by $N_C = C_t - A_t$, with a statistical error given by $E_{N_C} = \sqrt{\left(E_{C_t}\right)^2 + \left(E_{A_t}\right)^2 }$. One important metric commonly used to characterize the quantum correlation between photon pairs is the coincidence-to-accidental ratio (CAR), which is given by ${\rm CAR} = (C_t-A_t)/A_t$, with a statistical error of $\frac{E_{\rm CAR}}{\rm CAR} = \sqrt{\left(\frac{E_{C_t}}{C_t}\right)^2 + \left(\frac{E_{A_t}}{A_t}\right)^2 }$. The CAR values and pair generation rate were obtained with an integrating window of $\Delta \tau = \tau_{\rm FWHM}$, to include the major contribution of the coincidence spectrum (Fig.~\ref{Fig5}).

The pair generation rate was obtained by calibrating the true coincidence counts per gate with the clock frequency, duty cycle, and quantum efficiency of the detectors, as well as the transmittance from the device to the detectors for the photon pairs, given as
\begin{equation}
	R_{pair} = \dfrac{N_C}{(\eta_{t1} D_{t1})( \eta_{t2} D_{t2}) T \delta}, \label{R_pair}
\end{equation}
where $\eta_{t1}$ = 0.24 and $\eta_{t2}$ = 0.25 are the transmittance from the microresonator to the SPADs (including the tapered fiber, de-multiplexer, tunable bandpass filter, optical switch and beam splitter, see Fig.~\ref{Fig3}), $D_{ti}$ = 0.15 is the quantum efficiency of the SPAD, T is the total data acquisition time, $\delta$ is the duty cycle of the SPAD gating and i = \{1,2\} refers to the separate twin photon paths. The statistical error of the pair generation rate is obtained accordingly from $E_{N_C}$.

Figure \ref{Fig5}(b) plots the pair generation rate and CAR as a function of the total pump power (including both pumps) dropped into the cavity. We recorded a maximum CAR of 660 $\pm$ 62 with a pair generation rate of $(2.47 \pm 0.04) \times 10^4 ~{\rm pairs/s}$ at a dropped power of $P_d$ = 7.0 $\mu W$. Even with a small combined dropped power of $P_d$ = 37.6 $\mu W$, we were able to generate twin photons with a large pair generation rate of $(3.96 \pm 0.03) \times 10^5 ~{\rm pairs/s}$ and CAR of 95 $\pm$ 2. Note that these pair generation rates were obtained by integrating over $\tau_{\rm FWHM}$, which omits a significant number of true coincidences due to the full width at half maximum of the temporal correlation function, $\tau_{\rm c} = ( 2 \ln (2)) / \Gamma_t \approx 306$ ps (see Eq.~(\ref{Prob_pair})), being comparable to the instrument response time of the measurement system, $\tau_{\rm IRF} \approx 617$ ps. To obtain the total true coincidences, we increased the integration window. By varying the size of the coincidence window, we found that $R_{\rm pair}$ saturated at $\Delta \tau = 5 \times \tau_{\rm FWHM}$, which resulted in a maximum pair generation rate of $(5.29 \pm 0.04) \times 10^5 ~{\rm pairs/s}$. 

In summary, we have demonstrated the generation of twin photon pairs in a high-Q silicon microdisk resonator. We measured a record CAR of 660 $\pm$ 62 and pair generation rate as large as $(3.96 \pm 0.03) \times 10^5 ~{\rm pairs/s}$. The twin photon pairs were produced at telecommunication wavelengths using micro-Watt power levels on a CMOS compatible platform. As the field of quantum photonics makes rapid advances toward practical applications, the underpinning requirement of high fidelity quantum interference will remain.\cite{OBrien09,Guzik12,Spring13,Politi09,Knill00} High-purity single-mode twin photon pairs, which are identical in all degrees of freedom and exist within a narrow bandwidth, distinguish themselves as a great candidate for being the information carriers in such systems.

This work was supported by the National Science Foundation under Grant No. ECCS-1351697. It was performed in part at the Cornell NanoScale Science \& Technology Facility (CNF), a member of the National Nanotechnology Infrastructure Network, which is supported by the National Science Foundation.

\end{document}